\documentclass[slac_one]{revtex4}
\usepackage{graphicx}
\usepackage{fancyhdr}
\begin{document}

\title{{\small{Hadron Collider Physics Symposium (HCP2008), Galena, Illinois, USA}}\\ 
\vspace{12pt}
Electroweak Physics at LHC} 

\author{Kajari Mazumdar\\
On behalf of ATLAS and CMS Collaborations}
\affiliation{Tata Institute of Fundamental Research, Homi Bhabha Road, Mumbai 400 005, India.\\
}

\begin{abstract}
The Large Hadron Collider, LHC, though meant for discovery, will provide enough data from early phase to also perform various studies of Standard Model processes in as yet unexplored kinematic regions. Precision measurements of the electroweak variables will be possible due to the large rates of W and Z boson productions  combined with clean leptonic signatures.  Examples of simulation results from CMS and ATLAS collaboration studies are presented to show the wide variety of measurements possible and how various issues like background estimation, determination of systematic effects will be taken care of by the experiments. 
\end{abstract}

\maketitle

\thispagestyle{fancy}

\section{Introduction}
The Large Hadron Collider (LHC) machine will enter the collision mode at centre-of-mass energy of 10 TeV in 2008 and eventually will operate at 14 TeV from 2009 onwards gradually improving upon the luminosity. Two general-purpose detectors,  CMS (Compact Muon Solenoid) and ATLAS (A ToroidaL ApparatuS),  are ready for the data-taking phase with intense detector commisioning works underway. 
The much awaited discoveries at the LHC have to be preceeded by the {\it rediscovery} of the Standard Model (SM), and early discovery, though possible, will be challenging. On the other hand, from early phase on, {\it ie}, starting from an integrated luminosity of mere 1 pb$^{-1}$, various SM processes can be studied; in particular, heavy quarkonia peaks of $J/ \psi$ and $\Upsilon$ will shine well above background in the low mass range of the invariant mass distribution of dileptons. Electroweak (EW) processes of W and Z productions, having very large rates combined with clean leptonic decay mode signatures ($e, \mu$), are not only the {\it first physics}  at the LHC, they will be studied at all luminosities. These will be the {\it standard candles} for a large variety of LHC measurements. The W, Z physics  will enable, as a function of accumulated data, detector and physics-analysis tools commissioning of the experiments, precision determination of electroweak variables as well as measurements  in as yet unexplored kinematic regions. These programmes constitute both the confirmation as well as the consistency checks of SM at the highest energies LHC offer. The breakthrough discoveries at the LHC will be interpreted  by critically studying SM processes with experimental accuracy matched by theoretical understanding at the same level. 

Since the production cross section and dynamics are largely controlled by QCD, events with inclusive jets, photons, dileptons, heavy quarks will be produced copiously at the LHC, as exemplified in Table~\ref{evtrate}, providing complementary datasets which will eventually probe many aspects of strong and EW interactions with high precision. The large samples will also be useful for  search of rare processes which potentially hint at as yet unknown {\it New Physics}. 
\begin{table}[t]
\begin{center}
\caption{Expected triggered event yield in ATLAS with 1 pb$^{-1}$ data at 14 TeV.}
\begin{tabular}{|l|c|}
\hline \textbf{Channel} & \textbf{Number of Events}
\\
\hline
$\rm W \rightarrow \mu\nu$ & 7000\\
$\rm Z \rightarrow \mu\mu$ & 1100\\
${\rm t\bar t \rightarrow \mu +X}$ & 80\\
QCD jets with $p_T \ge 150$ GeV & 1000 (for 10\% trigger bandwidth)\\
Minimum bias & Trigger Limited \\
gluino-gluion($M\approx$ 1 TeV) & 1 - 10 \\
\hline
\end{tabular}
\label{evtrate}
\end{center}
\end{table}

 Precision calculations are mandatory for precision measurements. Large uncertainties due to soft gluon emission affect the predictions for transverse momentum, rapidity distributions of the bosons. Fortunately, accurate predictions for W, Z production, due to Next-to-Next-to-Leading-Order (NNLO), are available. Importantly, at the LHC energies the electroweak corrections to hadro-production of gauge bosons becomes significant and the K-factor for EW corrections at NLO is larger than that for QCD corrections at NNLO: ${\rm K_{EW} (NLO) } \ge {\rm K_{QCD} (NNLO)}$. Unfortunately no MC generator, which takes care of both, is available at the moment. W and Z inclusive cross-sections at parton level are calculated to a level of 3\% which can be compared with the uncertainty of 10\% for $t\bar t$ production process.

The Parton Density Function (PDF)s are crucial factors in rate estimation of any process at the LHC and the extrapolation from lower energy experiments results in large uncertainties contributing to the systematics of experimental study. The uncertainty in the theoretical estimate of the W and Z boson production due to PDFs affect the predictions for the transverse momentum and rapidity of the boson and the spectrum of the decay lepton as well. Various measurements based on W and Z processes including the differential distributions will be used in a consistent way to determine the structure funtions of the proton at the LHC energies. The parton $x$ range relevant for W production can be probed by measuremnets on Z.

While at tree tree level, the mass of the W boson is completely determined by 3 parameters of the electroweak theory, the radiative and loop corrections cause the mass to be sensitive to other particles due to the relation
\begin{equation}\label{eq:dr}
M_W = {\sqrt{\frac {\pi \alpha}{\sqrt 2 G_F}}}{\frac{1}{sin\theta _W {\sqrt{(1- \Delta R)}}}}
\end{equation}
where $\Delta R$ varies as $m_{top} ^2$, ${\rm log}M_H$ in SM. Precision measurements of the mass of the W boson allow the masses of presently undetected particles (such as the Higgs boson) to be inferred and to place constraints on new particles with weak charge via $\Delta R$.

Measurement of gauge boson pair productions, having reasonably high rate at the LHC, are also some of the first EW physics topics which will eventually lead to the determination of triple gauge couplings with unprecedented accuracies. The strength of the LHC is  that it makes sufficient number of Ws and Zs in the right kinematic region to explore the boson sector couplings ($WW\gamma, WWZ, ZZ\gamma$, etc.).

In this report we present some of the potential physics which can be done with accumulated data of few tens of  pb$^{-1}$ and a maximum of 1 fb$^{-1}$, {\it i.e.,} within a reasonably short time after the LHC start-up. The performance of ATLAS and CMS experiments are comparable and  not all the studies performed could be included and even the extent of coverage is limited in this short review. The  choice for discussion in this report is purely personal. Though experiments have prepared physics programme to be carried out with data of 10 TeV run, we confine here only to 14 TeV physics issues.

\section{DETECTORS AT LHC STARTUP} 
CMS and ATLAS detectors provide almost hermetic coverage around the collsion point with precision measurement subdetectors. As the names suggest the design of magnets are completely different but final resolutions and fiducial volumes are similar in both cases. W and Z processes  are essential from the earliest phase of data-taking for alignment, in-situ calibration, extraction of trigger and offline lepton identification efficiencies etc..  

At start up the  momentum scale is known to 1\% for muons which will eventually improve to a level of 0.02\%. The amount of material in front of electromagnetic calorimeter, to be determined by measureing E/p and photon conversions, if known to 1\%, will lead to a scale uncertainty of 0.01\%.
Similarly,  the final state radiation (FSR),  which affects the lepton identification requirements via isolation, if known to 10\% level, leads to a scale uncertainty of 0.01\%. The calibration of missing transverse energy will be performed with events of type $\rm W \rightarrow \mu\nu$ and  $\rm Z \rightarrow \tau\tau$.  Measurement of $\rm W \rightarrow \tau\nu$ cross-section is crucial for validation of $\tau$-identification algorithms and so on. Table~\ref{calib} gives some idea about the physics channels and the detector aspects they are crucial for.
\begin{table}[t]
\begin{center}
\caption{Data samples to be used for estimation of detector performance  at start-up}
\begin{tabular}{|l|c|c|c|}
\hline \textbf{Measurement} & \textbf{ATLAS} & \textbf{CMS} & \textbf{Physics process}
\\
\hline
ECAL uniformity & 1-2\% & 4\% & Isolated electrons, $\rm Z\rightarrow e^+e^-$\\
$e,\gamma$ energy scale & $\approx$ 2\% & $\approx$ 2\% & $\rm Z\rightarrow e^+e^-$\\
HCAL uniformity &$\approx$ 3\% & $\approx$ 2-3\% & single pions, QCD jets\\
Jet energy scale &$\le$ 10\% &$\le$ 10\% &$\gamma/Z$+ 1jet, $\rm W\rightarrow jj$ in ${\rm t\bar t}$  events\\
Track alignment& 10-200 $\mu$m in $R\phi$ & 20-200 $\mu$m in $R\phi$ & General tracks, isoolated $\mu$ in $\rm Z\rightarrow \mu\mu$\\
\hline
\end{tabular}
\label{calib}
\end{center}
\end{table}

The basic knowledge of QCD interaction at LHC being poor the predicted accuracy of  charge mulitiplicity in minimum bias events is only 50\%  and hence they must be estimated with priority for precision physics. Fortunately, it will be possible with data collected for only limited duration at $10^{31}$ cm$^{-2}$ s$^{-1}$ after startup. The underlying events populate the region transverse to the direction of leading  jet activity. The crucial observables for the soft-interaction physics are the density of the charged particles $\frac{dN}{d\eta d\phi}$, the charged hadron momentum spectrum etc. which will be studied using special tracking algorithms for soft tracks ($P_T\ge 150$ MeV)  developed by the experiments. These measurements are crucial for fine-tuning various parameters in the Monte Carlo generators and are also important ingradients for analysis tools of high $P_T$ physics like jet and lepton isolation, energy flow, jet tagging, etc. The startup data is of paramount importance to improve the inter-calibration of calorimeter modules by observation of azimuthal symmetry of deposited energy by minimum bias events.

\section{MEASUREMENT OF ELECTROWEAK PARAMETERS}
At the LHC the EW processes have such a large rate that statistical accuracy is not at all an issue.  $\sigma(pp\rightarrow  W \rightarrow \ell\nu) \sim$   20 nb and $\sigma(pp\rightarrow  Z \rightarrow \ell\ell) \sim$  2 nb, according to NNLO calculations. At low integrated luminosity of 1 fb$^{-1}$ the LHC experiments will be able to record about $10^7$ and $10^6$ W and Z  boson decays to electron and muon channels. Importantly, these processes have very little background and hence robust selections with loose invariant mass cuts will still allow high signal efficiency and high purity.  While the accuracy of the cross section extracted will be dominated by the integral luminosity measurement, ratios of W and Z productions will be important during early measurements from the LHC. During early measurement the uncertainty on luminosity will be be about 5\% which is expected to go down to about 1\% eventually with accumulated luminosity.

To derive precise measurements of the electroweak parameters, ie.,${\rm M_W}$, ${\rm \Gamma _W}$ and $sin^2 \theta$ the relevant observables are transverse momentum of the lepton, transverse mass of the W, ratio of distributions for W and Z, forward-backward asymmetry etc.. The asymmetries for the vector bosons productions (as well as in  their leptonic decay modes) will provide an independent, precise determination of the weak mixing angle, which will be competitive with the current world average. The forward-backward asymmetry in $\rm Z\rightarrow \ell\ell$ events is directly related to $sin^2\theta$. In this analysis the qurak direction is to be taken from the direction of the Z-boost. 
Since Z production is the highest around $|\eta| \sim 3$, measurement at high values of $y_{\ell\ell}$ is needed. This is experimentally possible only in electron  channel since electron acceptance extends to higher $\eta$ compared to muon acceptance.
For monitoring the collider luminosity or constraining the parton distribution functions (PDF), measurements of total cross-section, W-rapidity, W-charge asymmetry, lepton pseduorapidity etc. are important. It is to be noted that the EW corection at NLO for W-rapidity is of same order as the ammount of QCD ocrrection at NNLO and the uncertainty due to PDF. 

From experimental considerations, measurement of Z processes are always better than W processes, due to higher efficiency though having lower production rate. In Z-events both the decay products are measured accurately whereas in W-events, due to the presence of neutrino in the final state, the W-system is reconstructed only in the transverse plane of the beam direction. Further, the missing transverse energy is only moderate and there is no precise mass constraint. Thus the systematics are typically higher for W events combined with the presence of more background. Hence knowledge of various aspects, acquired from Z-events are applied to W-events.

\subsection{Early Measurements of W,Z in electron channel}
\begin{figure*}[t]
\centering
\includegraphics[width=80mm]{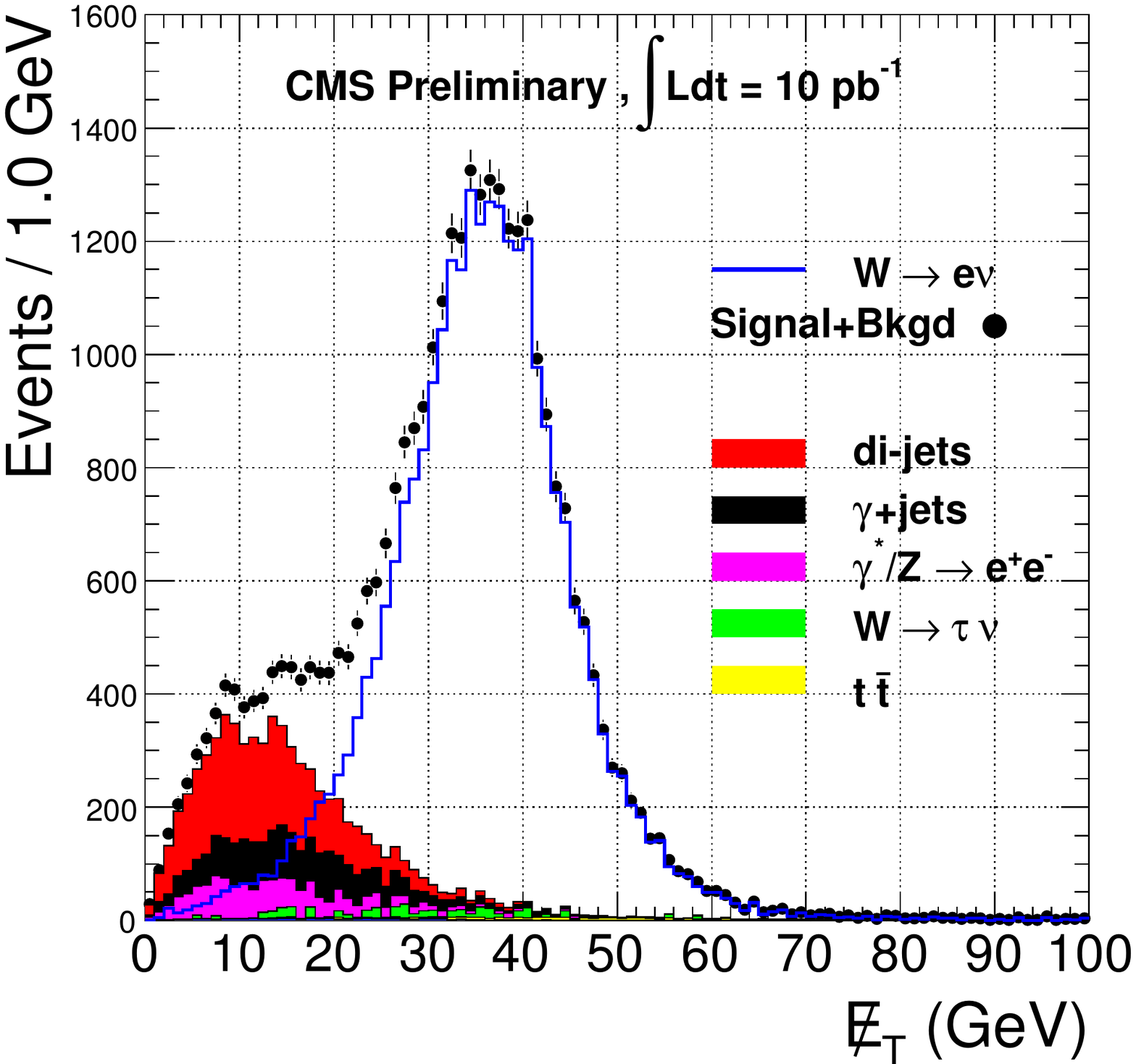}
\includegraphics[angle=90,width=80mm]{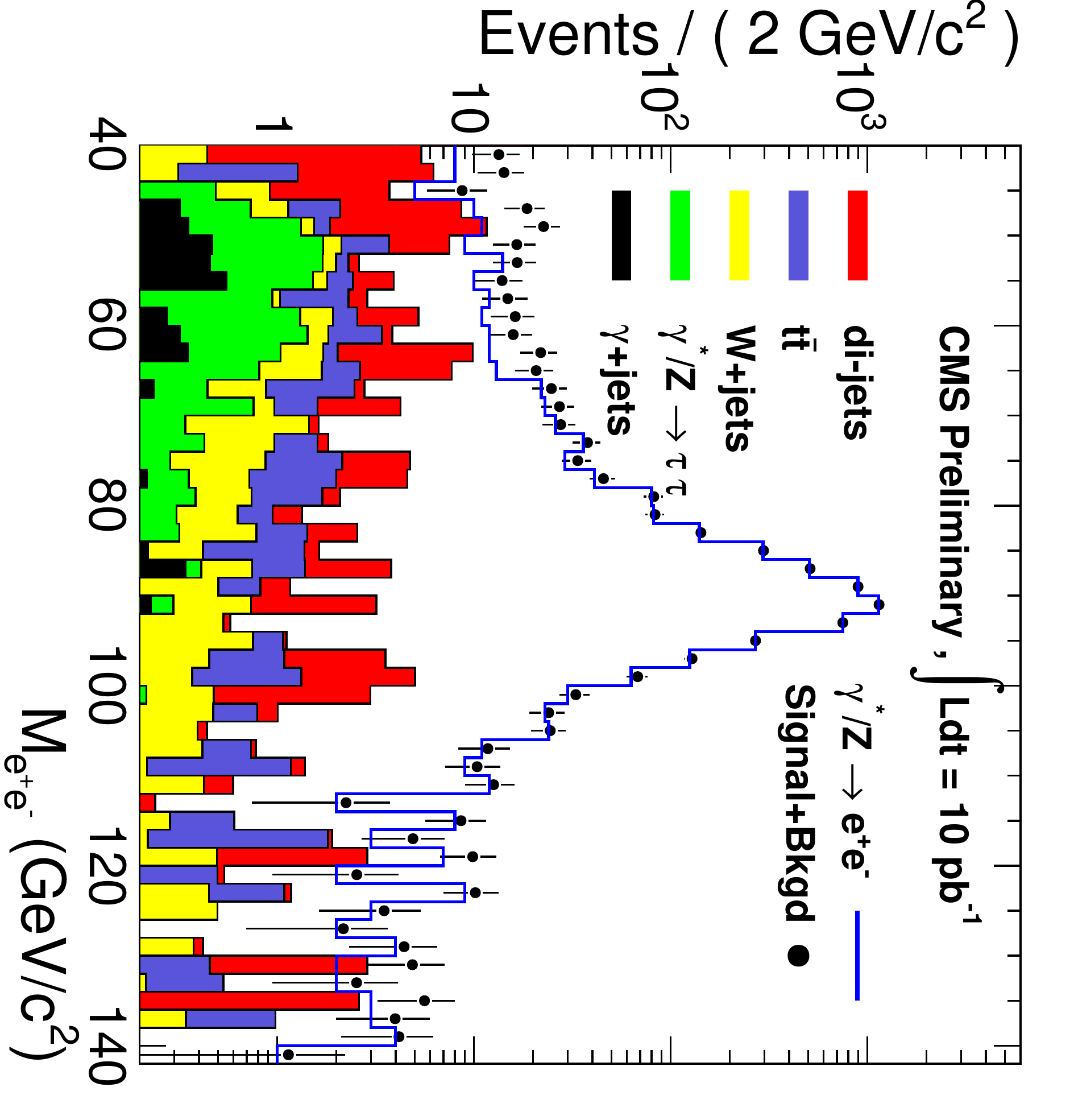}
\caption{Left: The missing transverse energy distribution for $\rm W \rightarrow e\nu$ events signal togther with considered backgrounds after selection cuts applied for an integrated luminosity of 10 pb$^{-1}$. Right: The invariant mass distribution of $e^+e^-$ pair for $\gamma ^*/Z \rightarrow e^+e^-$ signal events as well as backgrounds after selection cuts applied for an integrated luminosity of 10 pb$^{-1}$.} 
\label{el10pb}
\end{figure*}

Both ATLAS and  CMS plans to apply simple and robust criteria for electron selection during the initial stage of data taking when the calibration and alignment of the detector is imperfect.
 The missing $E_T$ spectrum measured in CMS for the $\rm W \rightarrow e\nu$ signal events alongwith various background contributions is shown in Fig.~\ref{el10pb}, left,  ~\cite{cmspase} corresponding to an integrated luminosiy of 10 pb$^{-1}$. The invariant mass distribution of oppposite-sign electron-pairs  is shown in Fig.~\ref{el10pb}, right, for $\gamma ^*/Z \rightarrow e^+e^-$ signal events as well as backgrounds. From data the selection efficiency (both offline and online) can be determined using Tag-and-Probe method. The signal and background rates used  correspond to the most accurate theoretical calculations and are obtained with full detector simulation taking into account the low luminosity conditions. In both cases the signal is highly prominent above SM background. The $\rm W \rightarrow e\nu$ cross-section is calculated using the formula (similar formula is used to determine $\gamma ^*/Z \rightarrow e^+e^-$ cross-section:\\
$$\sigma _W \times Br(W \rightarrow e\nu) ={\frac{N_W^{pass} - N_W^{bkg}}{A_W \times \epsilon _W \times {\cal L}}}$$
Here $N_W^{pass}$ is the number of candidate events, $N_W^{bkg}$ is the number of background events, $A_W$ is the acceptance defined as the fraction of decays which satisfy the geometric constraints of the detector and the kinematic constraints of the imposed selection criteria.  $\epsilon _W$ is the selection efficiency for W decays falling within the acceptance and ${\cal L}$ is the integrated luminosity.
Table~\ref{e10pb} gives the results expected from CMS experiment in the electron channel with 10 pb$^{-1}$ of data.
\begin{table}
\begin{center}
\caption{Results expected from CMS experiment in the electron channel with 10 pb$^{-1}$ of data.}
\begin{tabular}{|l|c|c|}
\hline 
\textbf{Quantity} & \textbf{$\rm W \rightarrow e\nu$} & \textbf{$\gamma ^*/Z \rightarrow e^+e^-$}\\
\hline
Nsel -Nbkg & 67954 $\pm$ 674 & 3914 $\pm$ 63 (Nbkg = 0)\\
\hline
Eff. for tag-and-probe (\%) & 65.1 $\pm$ 0.5 & 68.1 $\pm$ 0.6\\
\hline
Acceptance (\%) & 52.3 $\pm$ 0.2 & 32.39 $\pm$ 0.18 \\
\hline
Integrated Luminosity (pb $^{-1}$) & 10 & 10 \\
\hline
$\sigma \times$ Br (nb) & 19.97 $\pm$ 0.25 & 1.775$\pm$ 0.034\\
\hline
Cross-section used (nb) & 19.78 & 1.787 \\
\hline
\end{tabular}
\label{e10pb}
\end{center}
\end{table}
As we shall see the systematic uncertainties will start dominating for results obtained with data corresponding to an integrated luminosity above 1 fb$^{-1}$, the potential sources being: i) uncertainties in background estimate, ii) uncertainties in background subtraction/correction, iii) unaccounted correlations and various other sources of inefficiency. 

\subsection{Early measurements of W,Z in the muon channel}
With 50 pb$^{-1}$ data ATLAS experiment expects to have about 25.7k $Z\rightarrow \mu \mu$ events against approximately 100 background events. This will result in the measured cross-section of $\sigma = 2016 \pm 16$(stat) $\pm$ 64(syst)$\pm$ 202 (lumi) pb, using isolation in the inner tracker region of the detector. For $W\rightarrow \mu\nu$ events, using calorimetric muon isolation, the collected signal events will be about 300k against a background of about 20k and the measured cross-section is estimated to be $\sigma = 20530\pm 40$(stat) $\pm$ 630(syst)$\pm$ 2050 (lumi) pb. Hence, taking the ratio of the two cross-sections, the luminosity factor can be elimiated which will provide a stringent test of QCD.  

\begin{figure*}[t]
\centering
\includegraphics[width=75mm]{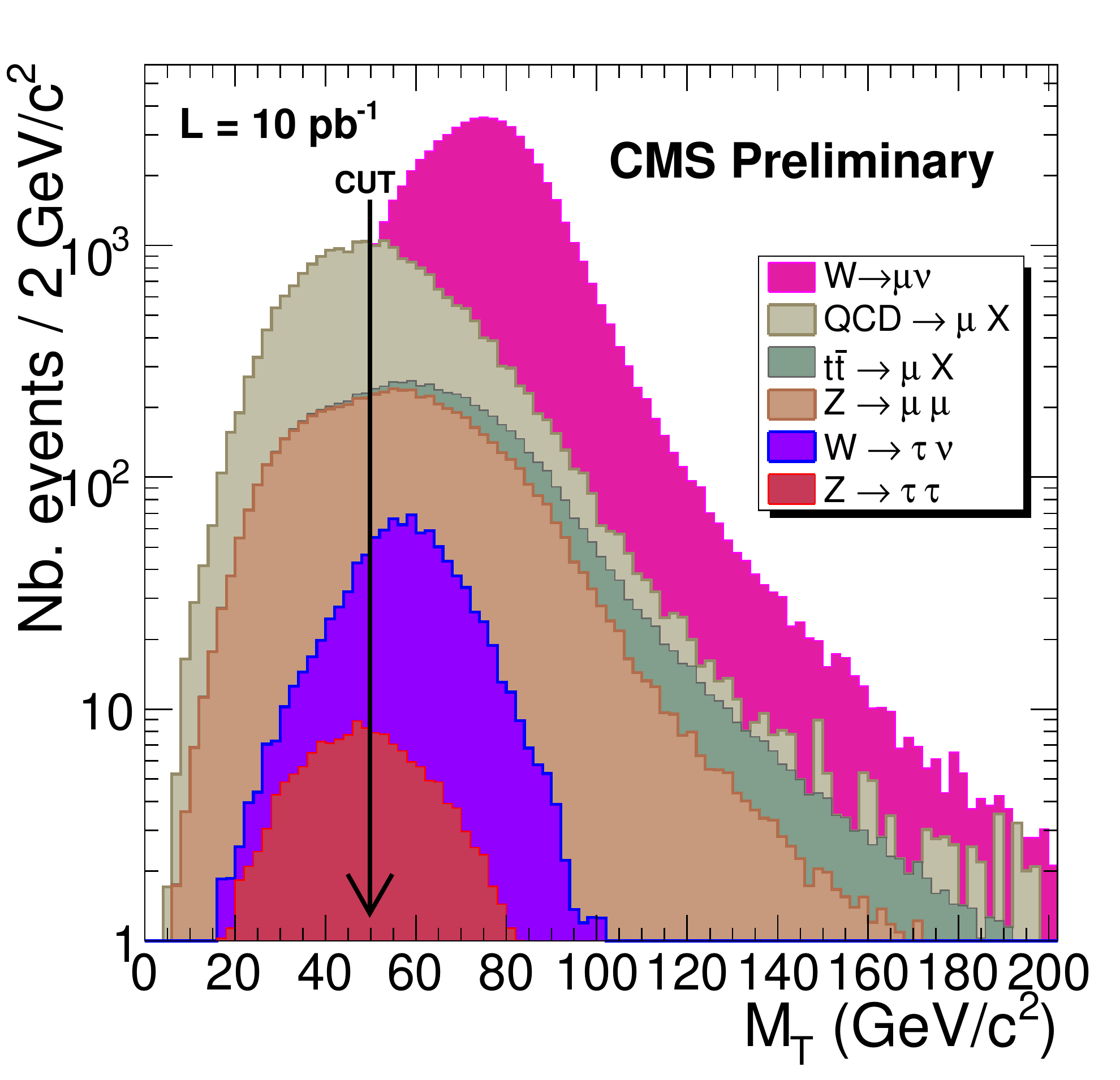}
\includegraphics[width=90mm]{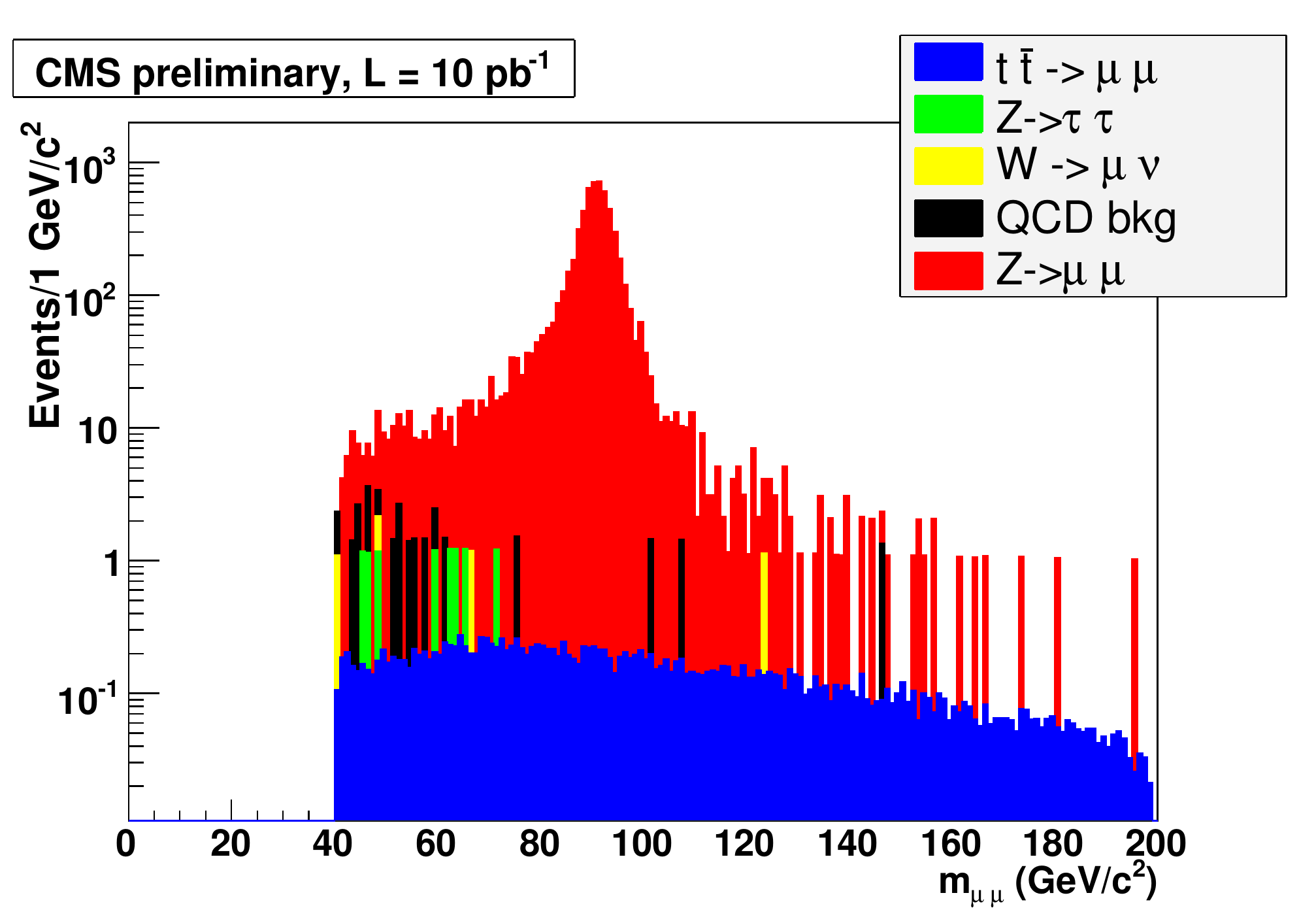}
\caption{Left: The reconstructed transverse mass ${\rm M_T}$ of W candidates in logarithmic scale for integrated luminosity of 10 pb$^{-1}$, including $\rm W \rightarrow \mu\nu$signal and the main backgrounds. Right: Selected $\rm Z\rightarrow \mu\mu$  candidates in logarithmic scale for integrated luminosity of 10 pb$^{-1}$, including $\rm W \rightarrow \mu\nu$signal and the main backgrounds. } 
\label{mu10pb}
\end{figure*}

CMS has developed detailed methods for the cross-section measurement for inclusive W and Z boson production in muon decay channel ~\cite{cmspasm}. 
Fig.~\ref{mu10pb}, left shows the reconstructed transverse mass ${\rm M_T}$ of $\rm W \rightarrow \mu\nu$ candidates in logarithmic scale. The predicted shapes are obtained from leading order generator packages like Pythia and Alpgen while the absolute normalization is determined, except for the case of QCD, from NLO calculations. The dominant background due to QCD is drastically reduced by accepting events with ${\rm M_T \ge}$ 50 GeV. Fig.~\ref{mu10pb}, right shows the invariant mass distribution of the selected di-muon candidates for the signal and background channels. 

\subsection{Determination of efficiency using data}
\begin{figure*}
\centering
\includegraphics[width=70mm]{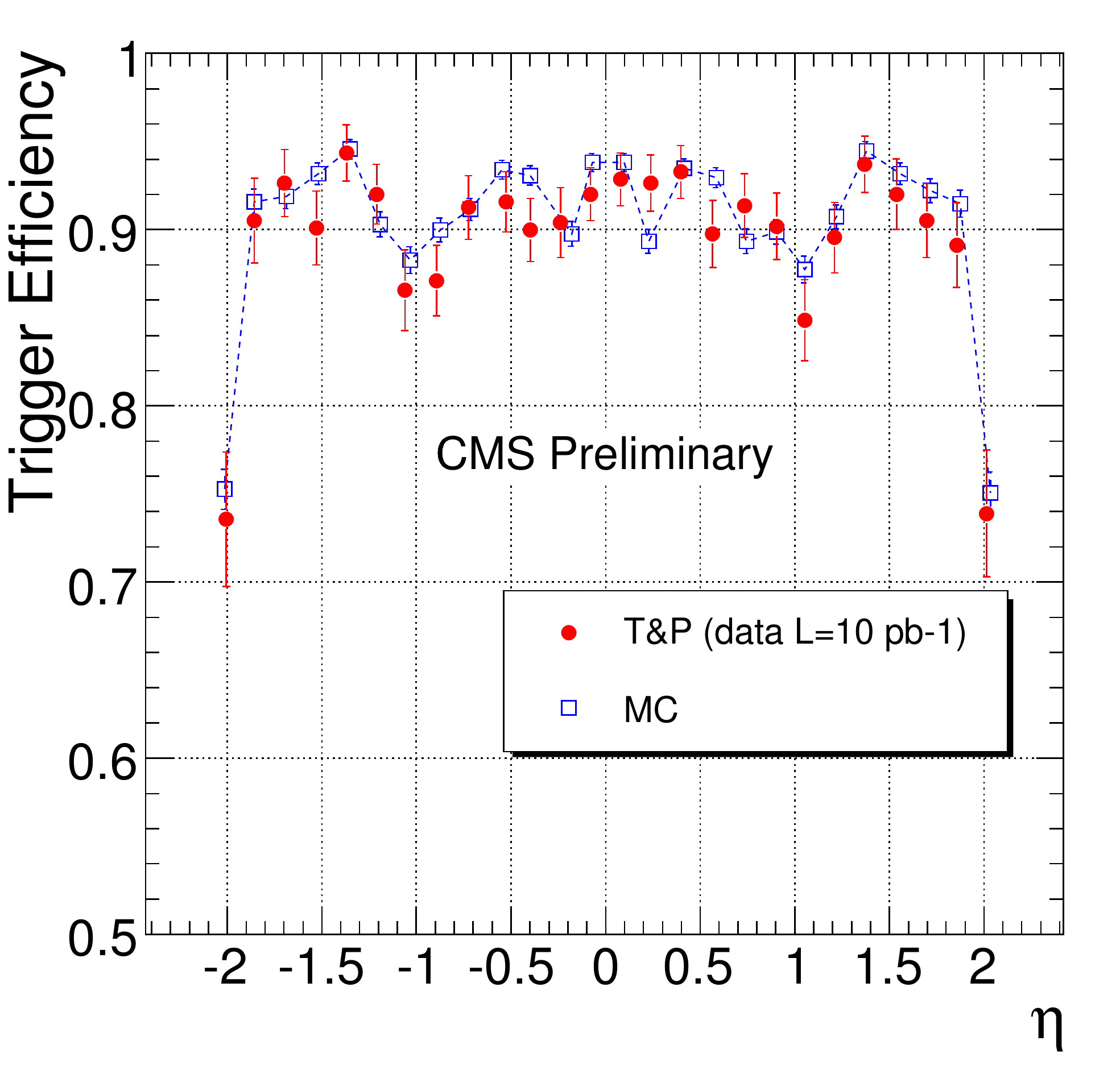}
\includegraphics[width=85mm]{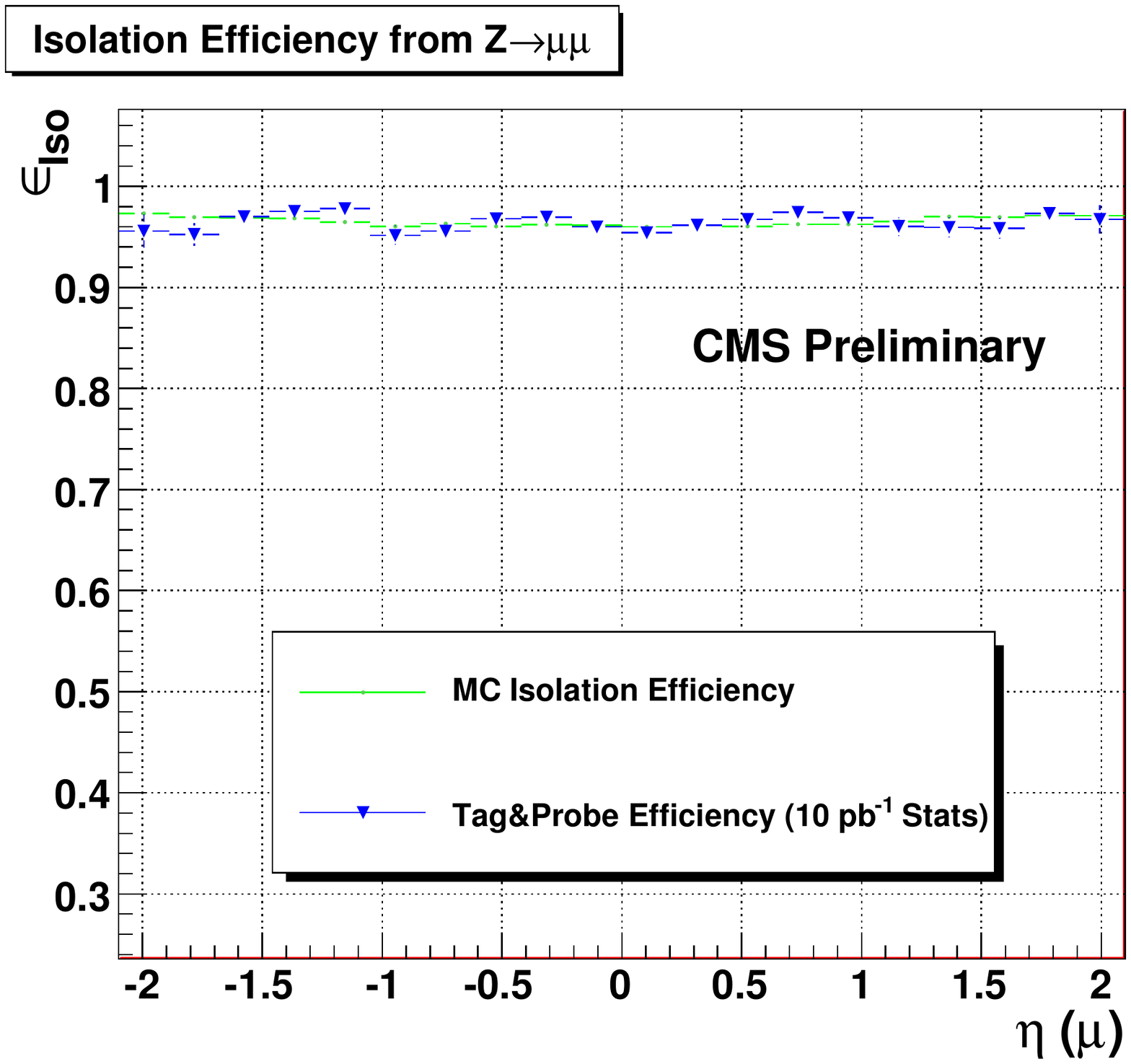}
\caption{Trigger and isolation efficiencies of muons in $\rm Z\rightarrow \mu\mu$ with $P_T \ge$ 20 GeV in CMS, determined by Tag-and-Probe method for 10 pb$^{-1}$ data, superposed with values determined from Monte Carlo information.} 
\label{mueff10pb}
\end{figure*}

During the early stage of the LHC experiment the dependence on yet-to-be-tuned Monte Carlo and the newly-commissioned physics tools will be reduced by determining various effeciency factors from the real data itself.
With data corresponding to 10 pb$^{-1}$, CMS plans to use a high purity sample of $\rm Z\rightarrow \mu\mu$ events where one muon is selected with strict quality criteria and is referred to as the {\it Tag}. The trigger, isolation and  selection efficiencies are measured directly on the other muon, referred to as the {\it Probe}. Fig.~\ref{mueff10pb}, left, shows trigger efficiency of high $P_T$ muons as a function of  pseudorapidity $\eta$ where values are obtained from both tag-and-probe method and generator level information. The right plot displays the isolation efficiency for muons with $P_T\ge$ 20 GeV as a function of $\eta$. The track isolation criteria typically requires the $P_T$ sum of all trackes in a $\Delta R = {\sqrt {\Delta \eta ^2 + \Delta\phi ^2}}$ cone of certain radius ({\it e.g., 0.3}) around the muon direction to be less than a threshold ({\it 3 GeV}). The results obtained from Tag-and-Probe  method is compared with the exact efficiencies obtained from Monte Carlo generator level information. In both cases the agreement is very good. 

\subsection{Estimation of background}
The most crucial background, due to QCD events, has to be determined from real data since the present knowledge of the rate has large theoretical uncertainty. For example, CMS plans to estimate the background for $\rm W \rightarrow \mu\nu$ channel using the so-called matrix method where 2 almost uncorrelated variables are analysed simultaneously which can discriminate the background against the signal, eg., $\Sigma P_T$ of tracks in muon isolation cone vs. reconstructed transverse mass of the W system, $M_T$. If 3 samples are selected as follows with statistics $N_B, N_C$ and $N_D$, the total number of QCD events is given by $N_{QCD} = N_B~N_C/N_D$ where\\
$\bullet$
$N_B$ = Number of events with nominal isolation cut $\Sigma _{tracks} P_T/P_T^\mu < 0.9$ and $M_T \le X$, where $X$ is optimised from Monte Carlo studies.\\
$\bullet$
$N_C$ = Number of events with nominal transverse mass cut $M_T \ge 50$ GeV and $\Sigma _{tracks} P_T/P_T^\mu \ge Y$ with $Y$ value optimised from Monte Carlo studies.\\
$\bullet$
$N_D$ = Number of events with  $M_T \le X$ and $\Sigma _{tracks} P_T/P_T^\mu \ge Y$.

The transverse mass shape for $\rm W \rightarrow \mu\nu$ events in the signal region can again be studied from data using missing transverse energy information as a function of $P_T$ of Z in  $\rm Z \rightarrow \mu\mu$ events. The template method for background evaluation uses a fit to the number of events for one of the backgrounds plus signal where signal shape is obtained in the first iteration from the signal $M_T$ distribution only. For determination of the background template the lepton isolation is reverted for the background enriched region of $M_T$ distribution.
\subsection{Improvement of $\rm Z\rightarrow \mu \mu$ cross-section with luminosity}
 With data corresponding to 100 pb$^{-1}$ the projected performances of the experiments, in terms of fractional errors in rates for  $\rm Z\rightarrow \mu \mu$ channel are:\\
$\bullet$ ATLAS: 0.0004 (stat) $\pm$ 0.008 (syst) $\pm$ 0.02 (th) $\pm$ 0.1(lumi)\\
$\bullet$ CMS:  0.0004 (stat) $\pm$ 0.011 (syst) $\pm$ 0.02 (th) $\pm$ 0.1(lumi)\\
Thus the statistical accuracy is already much lower than all other uncertainties and the dominant uncertainty is due to the luminosity error. The systematics of the measurements are at the level of 1\% dominated by efficiency measurement and background rate estimation etc. The theoretical error is due to the bias in the determination of acceptance and PDF uncertainties. With 1 fb$^{-1}$, experimental error due to uncertainty in determination of tracking and trigger efficiency is reduced to about 1 \%. The choice of PDF results in an
uncertainty of less than 1\% and the theoretical error due to higher order effects is the dominant one amounting to 2 to 3\%. CMS expects the cross-section values and their errors as\\
$\bullet$ $~\sigma (\rm W \rightarrow \mu\nu +X)$ = 14700 $\pm$ 6  (stat) $\pm$ 485  (syst) $\pm$ 1470 (lumi) pb \\
$\bullet$ $~\sigma (\rm Z \rightarrow \mu\mu +X)$ = 1160 $\pm$ 1.5  (stat) $\pm$ 27  (syst) $\pm$ 116 (lumi) pb  
 
\subsection{Drell-Yan events}
Drell Yan events are  theoretically the most clean physics process at hadron colliders, since the QCD effects enter in the initial state only. The production rate at the LHC being reasonably high, they will be studied from early stage, both below and above the Z-peak as a crucial confirmation of SM. They are also important background for Beyond SM physics searches as has been discussed in ~\cite{dimitri}.  
The differential distribution of the transverse momentum and the pseudo-rapidity leads to the determination of structure function. The $P_T^{\ell\ell}$ distribution, as a function of $M_{\ell\ell}$, will be very precise and can be utilised to infer $P_T$ distribution of W ~\cite{dy}. 

\subsection{Precision measurement of W mass $M_W$}
 The accurate determination of $M_W$ is a consistency check of SM. Also, even after the Higgs boson is discovered, this measurement will continue to be important since finding one Higgs is not necessarily the same as finding all of them, if they exist!
The LHC data will provide improved measurements of the W boson mass $M_W$ and the top quark mass $m_t$  which are crucial inputs for estimation of the Higgs boson mass $M_H$ due to the equation~\ref{eq:dr}. For equal contribution to  $\Delta M_H$,  $\Delta M_W$ should be $\approx 0.007 \Delta m_t$.  Thus for $\Delta m_t$ = 2 GeV, a value of $\Delta M_W$ = 15 MeV (per channel) will lead to an estimate of Higgs boson mass within  a range of 30\%. This demands that all contributions to W-mass uncertainty should be below 10 MeV, alternatively, for example, energy/momentum scale should be known to 0.02

Since the W decay products cannot be detected completely due to the escaping neutrino, transverse mass is reconstructed using the relation
$$M_T^W = \sqrt {2 P_T^\ell P_T^\nu (1-cos \Delta\phi)}$$
where the neutrino transverse momentum is reconstructed from the hadronic recoil $\vec P_T^h$ and the charged lepton system: $\vec P_T^\nu = -(\vec P_T^\ell + \vec P_T^h)$. The mass of the W is extracted by comparing the $M_T^W$ distribution observed from the data with that from samples simulated with varying $M_W$. But this method is affected by various instrumental and physical effects. Imperfect estimation of absolute energy scale due to uncertainties in kinematic distribution of W, non-gaussian tails of the energy distributions, recoil scale and resolution, the reconstruction efficiency are some of the experimental issues involved in improving $\Delta M_W$ . The theoretical sources of systematics are FSR which has direct effect on the lepton momentum and rapidity and transverse momentum distributions of W, which affect the W-width. Templates can be further affected by the environmnetal sources like backgrounds, underlying events, pile up, beam crossing angle etc. The situation thus underlines the need to devise alternative methods to determine $M_W$. 

Since the production model of W and Z bosons are same the QCD effects are almost similar in both cases. Thus Z-properties, which can be measured initially with better accuracies can be utilised to predict W-properties, like the spectrum of transverse momentum. Surely precision Monte Carlo is needed to account for the different phase space ($M_W \ne M_Z$) and different EW couplings have to be taken into account in the event distribution. Two approaches has been tried, the first one being  already used in Tevatron. By modelling $P_T^W$ with the measured $P_T^Z$, the estimated achievable precision in $\Delta M_W$ is about 20 MeV.

In the second method ~\cite{quast}, templates are created from Z-events. In the scaled observable method lepton $P_T$ distribution in $\rm W\rightarrow e \nu$ events are obtained in multiple steps.  $P_T^W$  is fitted with measured $P_T^Z$, where one lepton is randomly replaced by neutrino (or killed), the observable $X_V=P_T^e/M_V$ is rescaled and then the weight $R(X)$ is determined as
$$R(X) = \frac{d\sigma ^W /dX_W}{d\sigma ^Z /dX_Z}, X_V = P^e _T/M_V, V= W,Z$$
Then W selection is applied on the Z events with scale.
Most common uncertainties cancel and mainly $\Delta M_W$ is dominated by lepton energy scale linearity. The achievable error in 1 fb$^{-1}$ is 40 (stat.) + 40 (exp.) + 40 (theo.) MeV. In a slightly varied method called morphism, Z events are scaled instead of  scaling the observables. For $W\rightarrow \mu\nu$ events with 1 fb$^{-1}$ data, the morphism yields the errors as 40 (stat.) + 64 (exp.) + 20 (theo.) MeV, where the experimental error is dominated by the uncertainty on the scale of missing transverse energy
after correcting for the cross section ratio. The strong point about this method is that this distribution is exactly calculable in perturbative QCD and also soft gluon emission effects get cancelled in the ratio.

ATLAS has developed a method to calibrate the templates with constraints from Z events ~\cite{atlasmw}. Using very early data of 15 pb$^{-1}$ , the expected errors are   110 (stat.) + 114 (exp.) + 25 (PDF) MeV for electron channel using $P_T$ of the lepton and for muon channel, using $M_T$, the expected errors are 60 (stat.) + 230 (exp.) + 25 (PDF) MeV. With 10 fb$^{-1}$ data, ATLAS study anticipates to achieve $\Delta M_W \sim$ 6 MeV and with more luminosity and combination of experiments this error may go down to the level of 5 MeV. 

\subsection{Determination of proton structure function}
The kinematic regime at the LHC is much broader than currently explored and hence accurate knowledge of PDFs  is important for calculating the rates of SM physics,  which are backgrounds to any new physics discovery.
At the TeV scale, the uncertainties in cross section predictions for new physics are dominated by high-$x$ gluon uncertainty while at the EW scale (ie W and Z masses) theoretical predictions for the LHC are dominated by low-$x$ gluon uncertainty.  The rapidity $y$ is related with the structure funtion of the incoming partons $x_{1,2}$ through the relation $x_{1,2} = {\frac{M^2}{s}} e^{\pm y}$, where $\sqrt s =14$ TeV and $M$ is the mass involved in the reaction. Thus a central ($y=0$), heavy object of mass about 1 to 2 TeV has $x \sim 0.2$ which is also the relevant $x$ value in a Z boson event where $y_Z \sim 3.5$.

The $x$ dependence of the structure function $f(x,Q^2)$ is determined by fits to data, while the $Q^2$ dependence  is determined by the DGLAP equations.
For 1 fb$^{-1}$ of data it is expected that  $\rm W \rightarrow e\nu$ rapidity distribution will have experimental uncertainty sufficiently small to be sensitive to gluon shape parameter $\lambda$, where  $xg(x) \sim x^{-\lambda}$ and hence can distinguish between different PDF sets. An ATLAS study of rapidity distributions of electrons estimated an  improvement of about 35\% on the current value of $\lambda$, after ATLAS data was included in the global fit ~\cite{amanda}.

\section{DIBOSON PRODUCTION}
LHC will open up, for the first time, some of the di-boson ($V_1 V_2$) production channels. Experimentally, the invariant mass distribution of the diboson is studied to search for effects of {\it New Physics} which typically grow as $\sqrt s$ or $s$. If there is new physics at high $Q^2$ or short distances, it can manifest itself at lower $Q^2$ or longer distances as changes in the couplings. The guage structure of the underlying theory is determined by measuring these couplings accurately. For example, there are 14 possible $WW\gamma$ and $WWZ$ couplings, out of which about 5 are independent, CP conserving, EM gauge invariance preserving couplings. In SM,
$g_1^z =\kappa _\gamma = \kappa _Z =1$ and $\lambda _\gamma = \lambda _z = 0$ and conventionally $\Delta _g$,  $\Delta _\kappa$ and  $\Delta _\Lambda$  are measured keeping other couplings at their SM values. Angular distributions have additional resolving power since the W decays are self-analyzing while different couplings yield different angular distributions.
In the SM, there are no vertices containing neutral gauge couplings, like $ZZZ$ and $ZZ\gamma$ and at loop level the contribution is about $10^{-4}$.  Thus it is vital to measure these couplings at LHC though the improvements over current limits may have to wait for higher luminosity. NLO effects tend to increase the $V_1V_2$ cross-section at high $Q^2$  and this reduces the sensitivity towards anomalous couplings. 

\begin{figure*}[t]
\centering
\includegraphics[width=85mm]{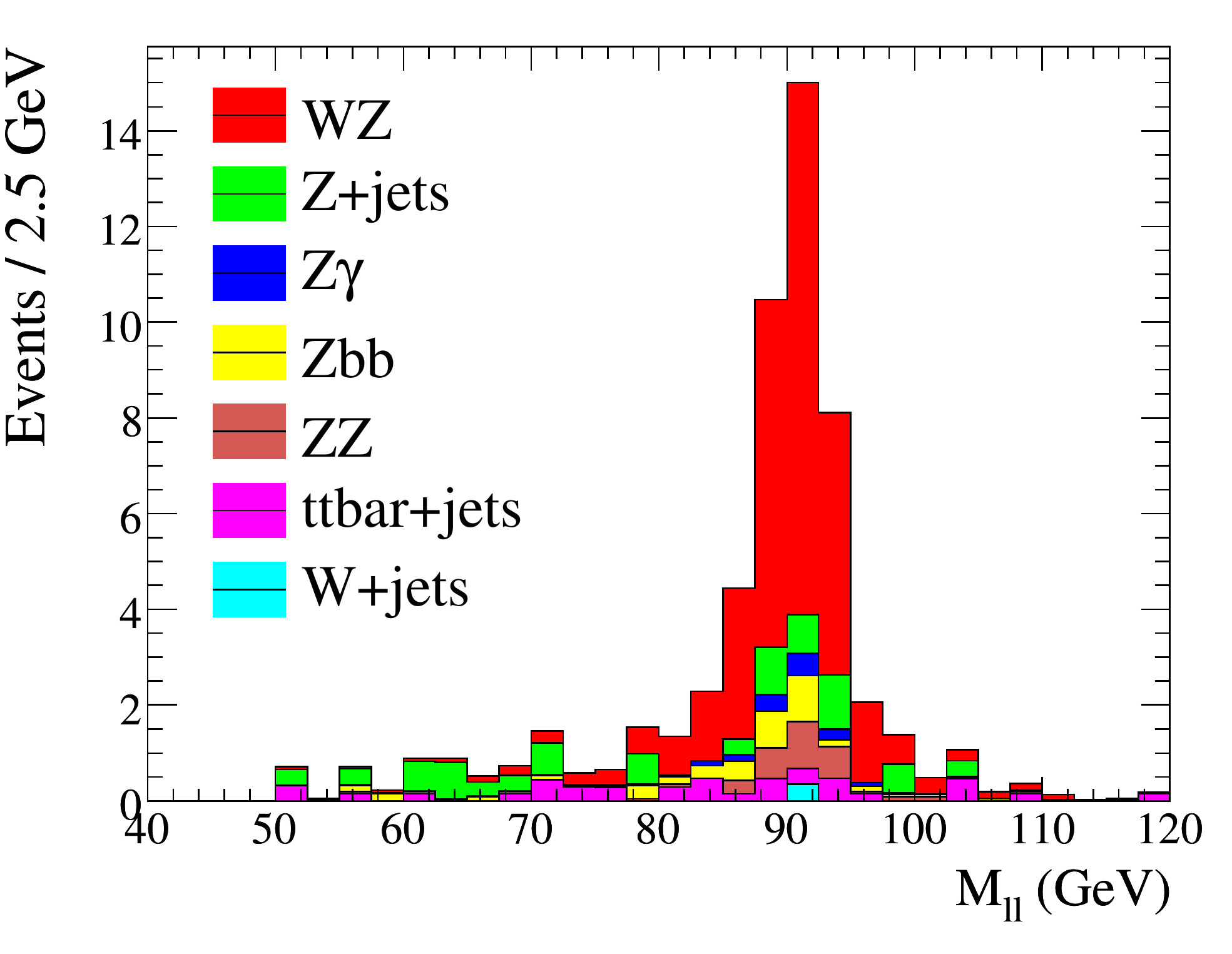}
\includegraphics[width=85mm]{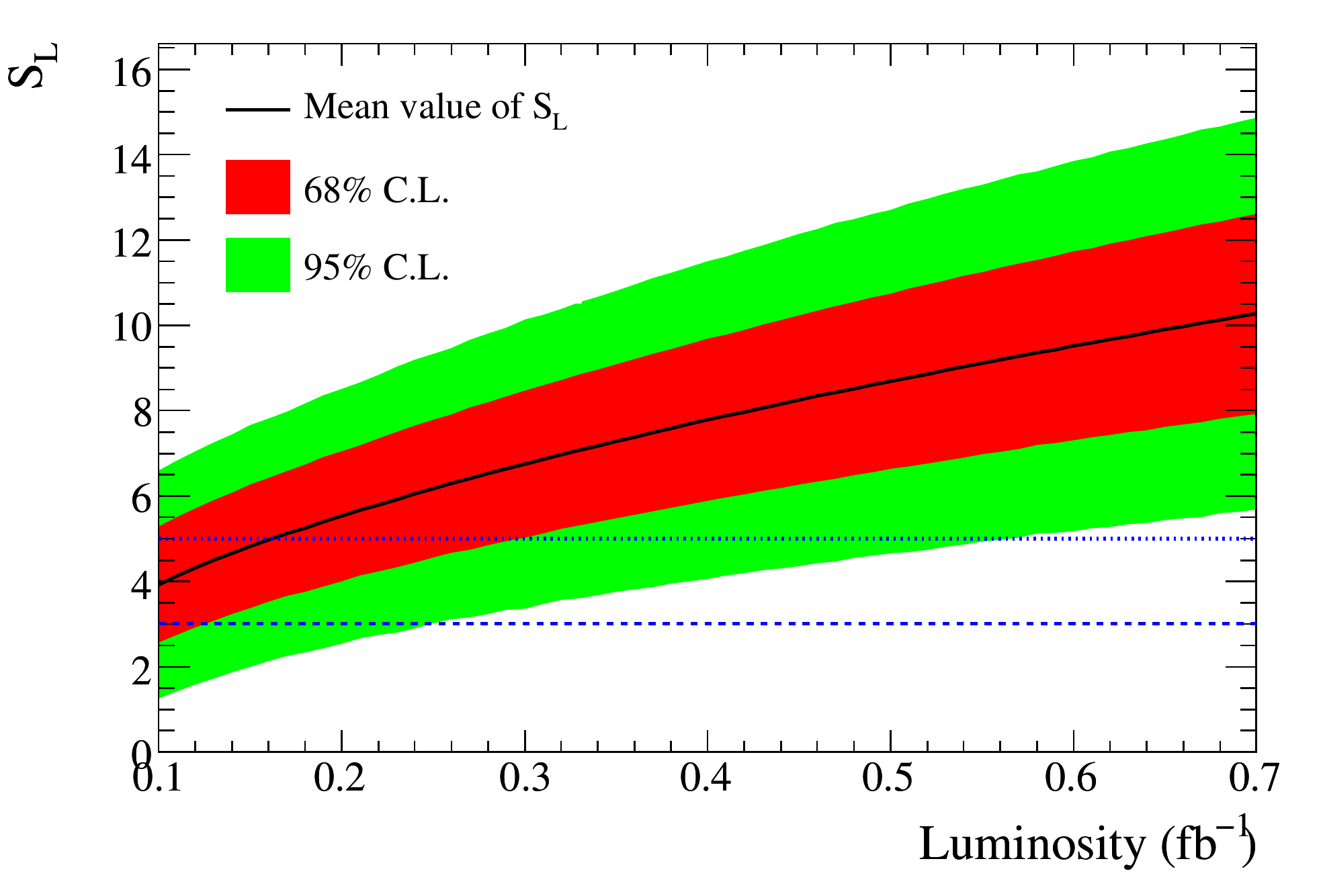}
\caption{Left: Invariant mass of $\rm Z\rightarrow \ell\ell$ candidate for all channels combined in CMS for 300 pb$^{-1}$. Right: Expected signal significance for ${\rm WZ}$ production as a function of integrated luminosity in CMS.} 
\label{wzresult}
\end{figure*}

Table~\ref{diboson} shows  the rates of some of di-boson production processes are high enough to be reacheable with data of $\le 1$ fb$^{-1}$. Moreover all these processes are genuine backgrounds for various searches, most notably for discovery of the Higgs boson and hence have to be measured accurately. With about 0.4 fb$^{-1}$ of data ${\rm ZZ} \rightarrow 4\mu$ process can be seen with a signal significance of about 5.
\begin{table}[t]
\begin{center}
\caption{Cross-section times branching ratio of various multiboson productions at LHC followed by leptonic decays.}
\begin{tabular}{|l|c|c|c|c|c|}
\hline \textbf{Channel} & ${\rm ZZ \rightarrow 4\ell (\ell = e,\mu)}$ &${\rm WZ \rightarrow 3\ell (\ell = e,\mu)}$ & ${\rm WW \rightarrow 2\ell (\ell = e,\mu)}$ & ${\rm WZ \rightarrow \ell \nu jj, \ell\ell jj}$ &${\rm W\gamma \rightarrow \ell\nu\gamma (\ell = e,\mu)}$ \\
\hline
\textbf{$\sigma _{\rm NLO} \times$ BR [pb]}& 0.037 & 1.1 & 12.0& 12 /2.3  & 400  \\
\hline
\end{tabular}
\label{diboson}
\end{center}
\end{table}
CMS expectation from a full simulation study of ${\rm WZ} \rightarrow 3\ell, \ell = e,\mu$ corresponding 300 pb$^{-1}$ ~\cite{cmspaswz} is shown in Fig.~\ref{wzresult}. Event yields and signal significance is given in Table~\ref{wzsyst1} where the systematics are also shown. Total systematic uncertainties in different channels will be different according to the type of lepton in the final state. Main sources of systematic uncertainties and their values  is shown in Table~\ref{wzsyst2}, other systematics like trigger etc., being at the level of 1 to 2\%. The signal significance is defined as $$S_L = \sqrt {2 {\rm ln} Q}, Q = (1+N_S/N_B)^{N_S +N_B} ~e^{-N_S}$$. A frequentist approach has been used to estimate the variation in the number of signal and background events.
\begin{table}[t]
\begin{center}
\caption{Event yields, signal significance and total systematic uncertainties for ${\rm WZ \rightarrow 3 \ell}$ events in CMS with 300 pb-1.}
\begin{tabular}{|l|c|c|c|c|c|}
\hline \textbf{Channels} & \textbf{no. of events} & \textbf{$S/{\sqrt B}$}& \textbf{Modelling} & \textbf{Background estimation, \%} & \textbf{Total, \%}\\
\hline
$3e$ & 7.9 $\pm$ 0.3 & 3.3 & 21 & 27 & 34 \\
$2e1\mu$ & 8.0 $\pm$ 0.3 & 7.2& 19 & 16 & 25\\
$2\mu 1e$& 8.9 $\pm$ 0.3 & 3.9 & 17 & 31 & 35\\
$3\mu$& 10.1$\pm$0.3 & 8.3 & 17 & 12 & 21 \\
\hline
\end{tabular}
\label{wzsyst1}
\end{center}
\end{table}
\begin{table}[t]
\begin{center}
\caption{Main sources of systematic uncertainties for ${\rm WZ \rightarrow 3 \ell}$ events in CMS with 300 pb$^{-1}$.}
\begin{tabular}{|l|c|c|c|c|}
\hline \textbf{Source} & Luminosity & ${\rm M_T (W)}$ requirement & PDF uncertainties & Electron identification\\
\hline
\textbf{Systemtaic uncertainty, \%}  & 10.0 & 10.0 & 4.0 & 4.0 \\
\hline
\end{tabular}
\label{wzsyst2}
\end{center}
\end{table}

\section{SUMMARY}
To establish the anticipated discoveries at the LHC, the Standard Model processes must be well understood.
Various aspects of Z-boson production and subsequent decays will be utilised by experiments for
precision determination of electroweak parameters.
An early, competitive W Mass measurement from the LHC is improbable.
Some of the di-boson productions are likely to be etsablished during the early phase of the LHC.

\begin{acknowledgments}
The author wishes to thank the organisers for the kind support and local hospitality without which the participation would not have been possible. The author also gratefully acknowledges many collaborators in CMS experiment for overall support in presenting the talk and preparing this report.
\end{acknowledgments}

\end{document}